\documentclass[aps,prl,twocolumn,groupedaddress,showpacs]{revtex4}

\usepackage{graphicx}
\usepackage{color}

\begin{document}


\title{Cooling Atoms with a Moving One-Way Barrier}


\author{Elizabeth A. Schoene, Jeremy J. Thorn, and Daniel A. Steck}
\affiliation{Oregon Center for Optics and Department of Physics, 1274 University
of Oregon, Eugene, OR 97403-1274}


\date{\today}

\newcommand{\change}[1]{%
	{\color{blue}%
	\ensuremath{\clubsuit\!\triangleright}#1%
	\ensuremath{\triangleleft\!\clubsuit}}%
}

\begin{abstract}

We implement and demonstrate the effectiveness of a cooling scheme using a moving, all-optical, one-way barrier to cool a sample of $^{87}$Rb atoms, achieving nearly a factor of 2 reduction in temperature.  The one-way barrier, composed of two focused, Gaussian laser beams, allows atoms incident on one side to transmit, while reflecting atoms incident on the other.  The one-way barrier is adiabatically swept through a sample of atoms contained in a far-off-resonant, single-beam, optical dipole trap that forms a nearly harmonic trapping potential. As the barrier moves longitudinally through the potential, atoms become trapped to one side of the barrier with reduced kinetic energy.  The adiabatic translation of the barrier leaves the atoms at the bottom of the trapping potential, only minimally increasing their kinetic energy.
\end{abstract}



\maketitle


The field of atom optics owes much of its success to the development of robust, highly effective laser-cooling techniques, though they are only applicable to a fraction of atomic species \cite{metcalf1999}.  Though duly celebrated, these techniques generally rely on the existence of a cycling optical transition---atoms falling into ``dark states'' decouple from the cooling lasers.  With a few exceptions \cite{yabuzaki99,celotta00,hanssen06,monroe07,katori08,lev10}, only the simplest of atoms meet this requirement, creating a need for new, innovative cooling tools to open the field of ultra-cold physics research to more complex atoms and molecules.  Several proposed techniques employ one-way barriers \cite{raizen2005,ruschhaupt2004,dudarev2005,kim2005,ruschhaupt2006,ruschhaupt2006b,ruschhaupt2006c,price2007,ruschhaupt2007,ruschhaupt2008,price2008}, where atoms ``see'' a repulsive or attractive optical potential depending on the side of incidence.  

The cooling utility of a one-way barrier is best understood by its relationship to Maxwell's demon \cite{ruschhaupt2006c,thorn2008,thorn2009}, an imaginary creature that controls a trap door partitioning a container of gas in a variation of Maxwell's famous thought experiment \cite{maxwell1871,bennett1987,scully2007}.  The demon uses the trap door to collect all the gas on one side of the container, reducing the volume occupied by the gas without increasing its temperature.  The one-way barrier functions in the same way as the demon and its trap door, allowing atoms traveling in only one direction to pass through, compressing the volume of an atomic sample confined in a trap.  Though Maxwell's demon appears to compress the gas without performing any work, this apparent violation of the second law of thermodynamics has many resolutions \cite{bennett1987,scully2007}, and one-way barriers are no exception \cite{thorn2009}.  In our experiment, an effective position measurement of each transmitting atom occurs as it spontaneously scatters barrier photons, which carry away the necessary entropy.  Phase-space compression of $^{87}$Rb has been achieved with one-way barriers \cite{price2008,thorn2008}, with the most recent work demonstrating a factor of 350 compression from the initial conditions \cite{bannerman2009}.  Cooling the compressed atomic sample can then be accomplished, in principle, via adiabatic expansion.

\begin{figure}
\includegraphics[height=2.5in]{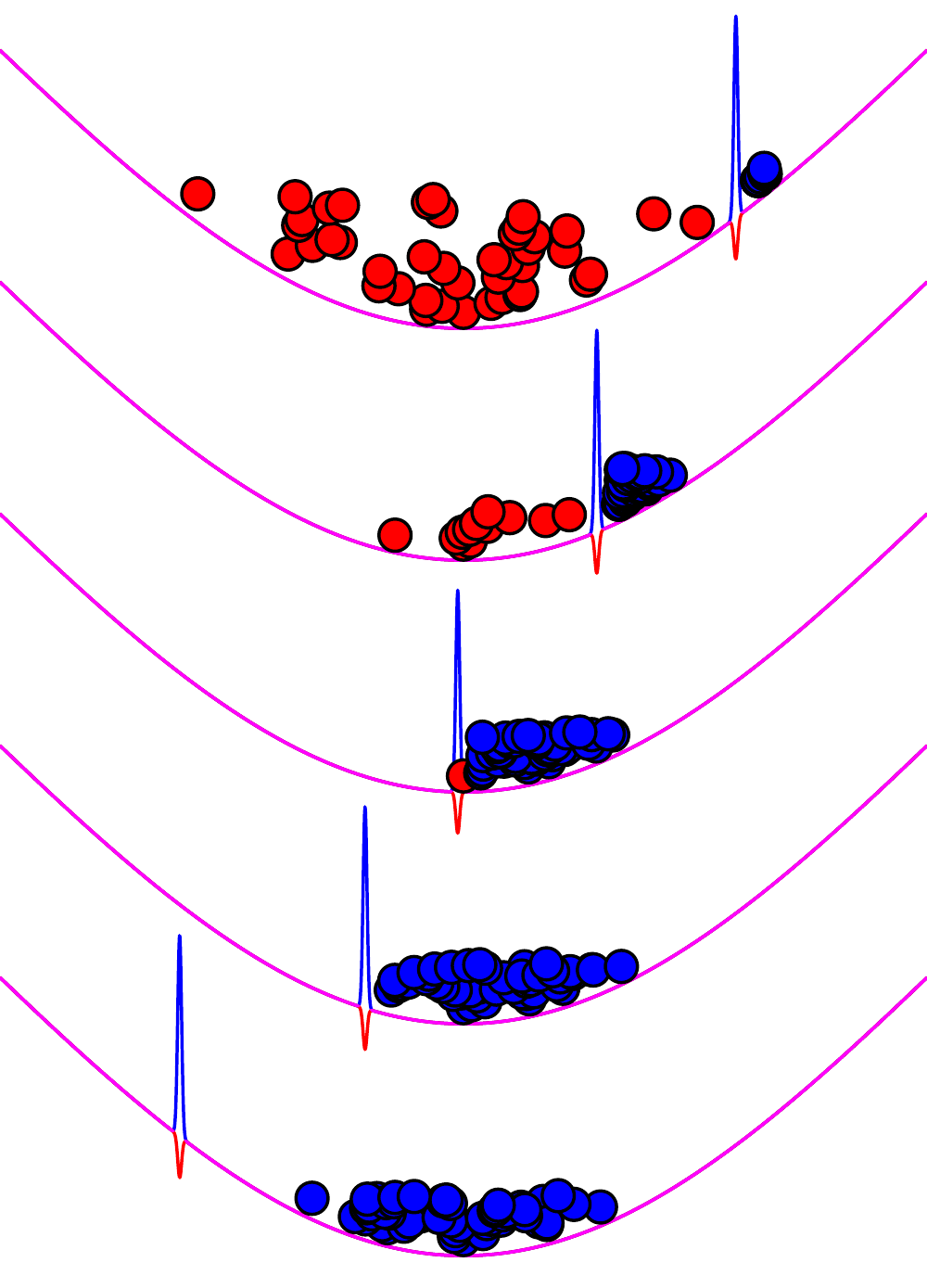}
\caption{(Color online.) Schematic representation of the one-way barrier cooling process for atoms trapped in a harmonic potential.  The atoms pass through the barrier from left to right and become trapped near their turning points, then reduce their potential energy by following the barrier to the bottom of the potential.\label{fig:cooling}}
\end{figure}

While intrinsically fascinating as a physical realization of Maxwell's demon, the real promise of one-way-barrier cooling schemes lies with the small number of scattering events necessary to achieve cooling.  In principle, only a single scattering event is required, bypassing complications due to dark states and the complex electronic structure of many atoms and molecules [\onlinecite{narevicius2009}].  Additionally, if the atoms reside in a harmonic trap, the compression and adiabatic expansion steps of the cooling process can be combined by adiabatically sweeping the barrier through the atomic sample (Fig.~\ref{fig:cooling}) [\onlinecite{ruschhaupt2006c}].  In this scheme, the oscillating atoms pass through the one-way barrier near their turning points, becoming trapped when they have very little kinetic energy.  The trapped atoms then follow the barrier to the trap center, lowering their potential energy with little increase in kinetic energy.

In this paper we investigate a particular implementation of the cooling technique described above, using a moving one-way barrier to cool atoms confined in a dipole trap.  We demonstrate the effectiveness of this technique at reducing the temperature of the atomic sample, and discuss limitations resulting from constraints in our experimental design. 



\begin{figure}
\includegraphics[width=\columnwidth]{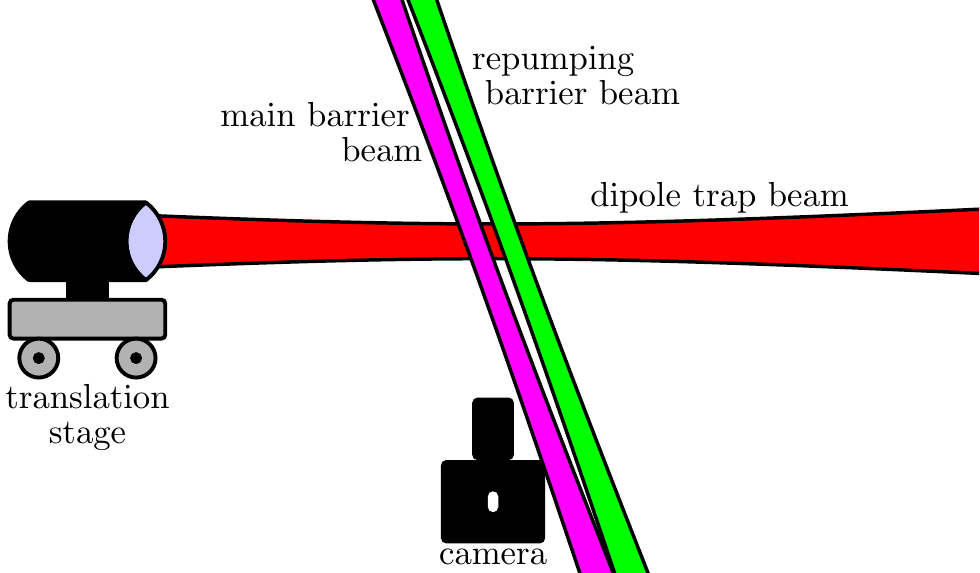}
\caption{(Color online.) Optical setup showing the dipole trap, barrier beams, imaging system, and translation stage.\label{fig:opticalsetup}}
\end{figure}

The essential component of the cooling scheme is the optical one-way barrier, whose interaction with cold atoms we have studied in detail, and for which the bulk of the experimental design and procedures remains the same [\onlinecite{thorn2008,thorn2009}].  The one-way barrier consists of two focused, Gaussian laser beams (Fig.~\ref{fig:opticalsetup}), with the main barrier beam tuned between the $F=1 \rightarrow F'$ and the $F=2 \rightarrow F'$ $^{87}$Rb hyperfine transitions, while the repumping barrier beam is resonant with the $F=1 \rightarrow F'$ repump transition.  Tuning the main barrier beam frequency $\omega$ between the two ground state resonances $\omega_0$, produces opposite signed detunings $\Delta:=\omega-\omega_0$ for the two hyperfine ground levels.  This results in the optical dipole potential due to the main barrier beam, which is inversely proportional to the detuning, presenting an attractive potential to atoms in the $F=1$ ground state, and a repulsive potential to atoms in the $F=2$ ground state.  Once atoms in the $F=1$ ground state pass through the main barrier beam, the repumping barrier beam pumps them to the $F=2$ ground state, thus flipping the atoms from the transmitting state to the reflecting state.

We use a double-magneto-optical-trap (double-MOT) system to initially cool and trap the $^{87}$Rb atoms in ultra-high vacuum, resulting in about $1.4 \times 10^{5}$ atoms at about $30 \ \mathrm{\mu K}$.  Next we load the atoms into a far-detuned, optical dipole trap formed by a Yb:fiber laser with a wavelength of $1090(5) \ \mathrm{nm}$, which we focus to a $30.9(5) \ \mathrm{\mu m}$ waist ($1/e^2$ intensity radius).  We operate the laser at $9.3(5) \ \mathrm{W}$, producing a nearly conservative, nearly harmonic potential well with a small-amplitude oscillation frequency of $24 \ \mathrm{Hz}$ in the longitudinal direction.  This potential has a maximum depth and scattering rate of $k_{_\mathrm{B}} \cdot 0.9 \ \mathrm{mK}$ and $3 \ \mathrm{s^{-1}}$, respectively, for $^{87}$Rb atoms in either hyperfine ground level.

We load atoms from the MOT into the dipole trap for $110 \ \mathrm{ms}$.  Magnetic bias fields allow us to load the atoms $0.95(5) \ \mathrm{mm}$ from the trap focus, which effectively increases the temperature of the dipole-trapped atoms.  Heating the sample ensures that we have the initial temperature well above the cooling limit for this scheme.  We measure a longitudinal oscillation period of $50 \ \mathrm{ms}$, which is longer than the harmonic frequency suggests.  The difference results from loading the atoms in the mildly anharmonic region of the trapping potential and from angular momentum, which can significantly reduce the longitudinal speed of the atoms \cite{thorn2009}.  This process fills up the trap, loading approximately $8 \times 10^{4}$ atoms, with a temperature of about $110 \ \mathrm{\mu K}$.  The atoms are then optically pumped to the $F=1$ ground state with a $15 \ \mathrm{ms}$ pulse of MOT trapping light.  We heat the atoms further using parametric excitation \cite{balik2009}, modulating the dipole-trap laser intensity with a square wave switching between $100\%$ and $120\%$ of the nominal operating power, and a period of $24 \ \mathrm{ms}$, for $100$ periods.  This results in approximately $4 \times 10^{4}$ atoms in the dipole trap with a temperature of about $170 \ \mathrm{\mu K}$.

The fiber collimator for the dipole-trap laser is mounted on a precision translation stage, enabling us to translate the trapping potential through the barrier (Fig.~\ref{fig:opticalsetup}).  This is technically simpler than translating the two barrier beams through the trap.  The two configurations are equivalent because the translation is done adiabatically, except the atoms experience a force due to the acceleration and deceleration of the trapping potential.  We translate the trap over a total distance of $4 \ \mathrm{mm}$ at a velocity of $5 \ \mathrm{mm/s}$ during the cooling sequence.

\begin{figure}
\includegraphics[width=\columnwidth]{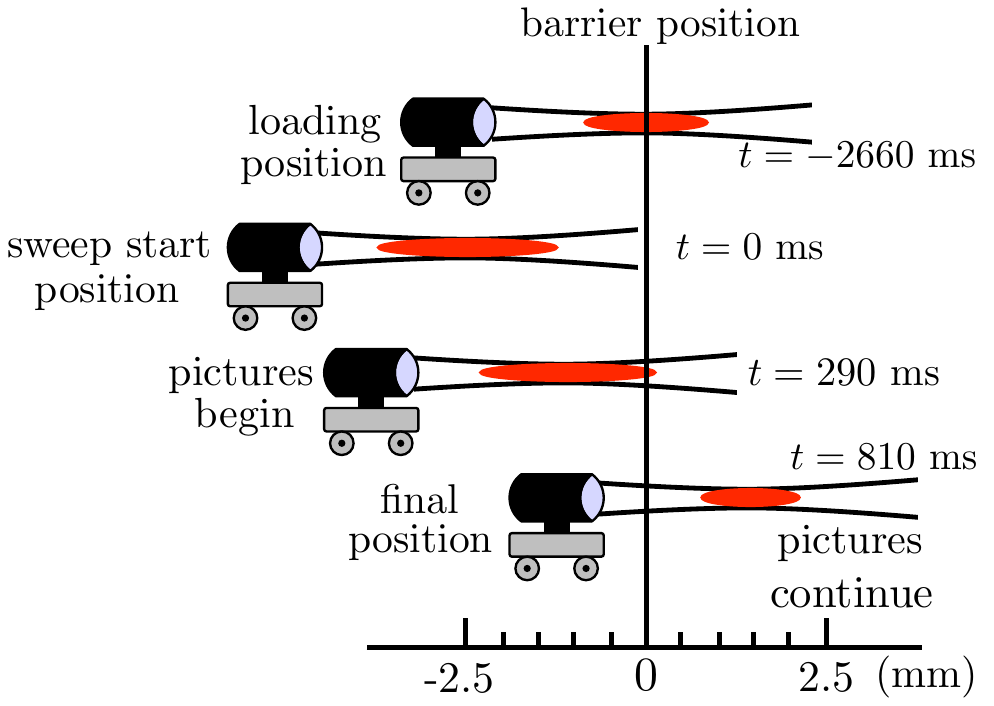}
\caption{(Color online.) Dipole-trap translation and imaging sequence.  The times shown correspond to a translation speed of $5 \ \mathrm{mm/s}$.\label{fig:movementsequence}}
\end{figure}

The geometry of the one-way barrier beams remains the same as described previously \cite{thorn2008,thorn2009}, except the beam separation is $36(1) \ \mathrm{\mu m}$, a separation that is optimized for minimizing heating on the reflection side \cite{thorn2009}.  We focus the main (repumping) barrier beam to a $11.5(5) \ \mathrm{\mu m}$ [$13(2) \ \mathrm{\mu m}$] waist along the dipole trap axis, and a $80(7) \ \mathrm{\mu m}$ [$60(7) \ \mathrm{\mu m}$] waist perpendicular to the dipole-trap axis, with a power of $24(2) \ \mathrm{\mu W}$ [$0.34(3) \ \mathrm{\mu W}$] inside the vacuum chamber.  The repumping barrier beam is resonant with the $^{87}$Rb $F=1\rightarrow F'$ repump transition, while the main barrier beam is stabilized to the $^{85}$Rb $F=3 \rightarrow F'=3,4$ crossover transition in the saturated absorption spectrum.  This produces a detuning $1.05(5) \ \mathrm{GHz}$ to the blue side of the $^{87}$Rb MOT trapping transition.

To image the atoms we illuminate them with a $45 \ \mathrm{\mu s}$ pulse of probe light resonant with the $F=2 \rightarrow F'=3$ MOT trapping transition, oriented transversely to the dipole-trap axis.  A charge-coupled-device (CCD) camera collects any resonant light not scattered by the atoms, imaging the shadow of the atom distribution.  Background offsets (computed from the edge regions of the images) are subtracted on a per-column basis to reduce systematic errors due to interference fringes in the images.

The sequence of events to produce a single image is as follows: At the location where the dipole-trap focus intersects the main barrier beam, the atoms are loaded into the dipole trap, then optically pumped to the $F=1$ transmitting state.  The trap is then translated $2.5 \ \mathrm{mm}$ to left of the barrier position (as seen by the camera) at $5 \ \mathrm{mm/s}$, while the atoms are simultaneously heated via dipole-trap intensity modulation for $2400 \ \mathrm{ms}$.  After heating, the barrier beams are turned on, and at time $t=0$ the trap is translated $4 \ \mathrm{mm}$ at $5 \ \mathrm{mm/s}$, to a position $1.5 \ \mathrm{mm}$ to the right of the barrier position.  As the atomic sample starts to pass through the barrier, we begin recording images every $30 \ \mathrm{ms}$ until about $500 \ \mathrm{ms}$ after the sweep has finished and the trapping potential has come to rest (Fig.~\ref{fig:movementsequence}).  It is important to note that many images are taken while the atoms are still in motion.  Since each measurement is destructive, the sequence is repeated for each image.



\begin{figure}
\includegraphics[width=\columnwidth]{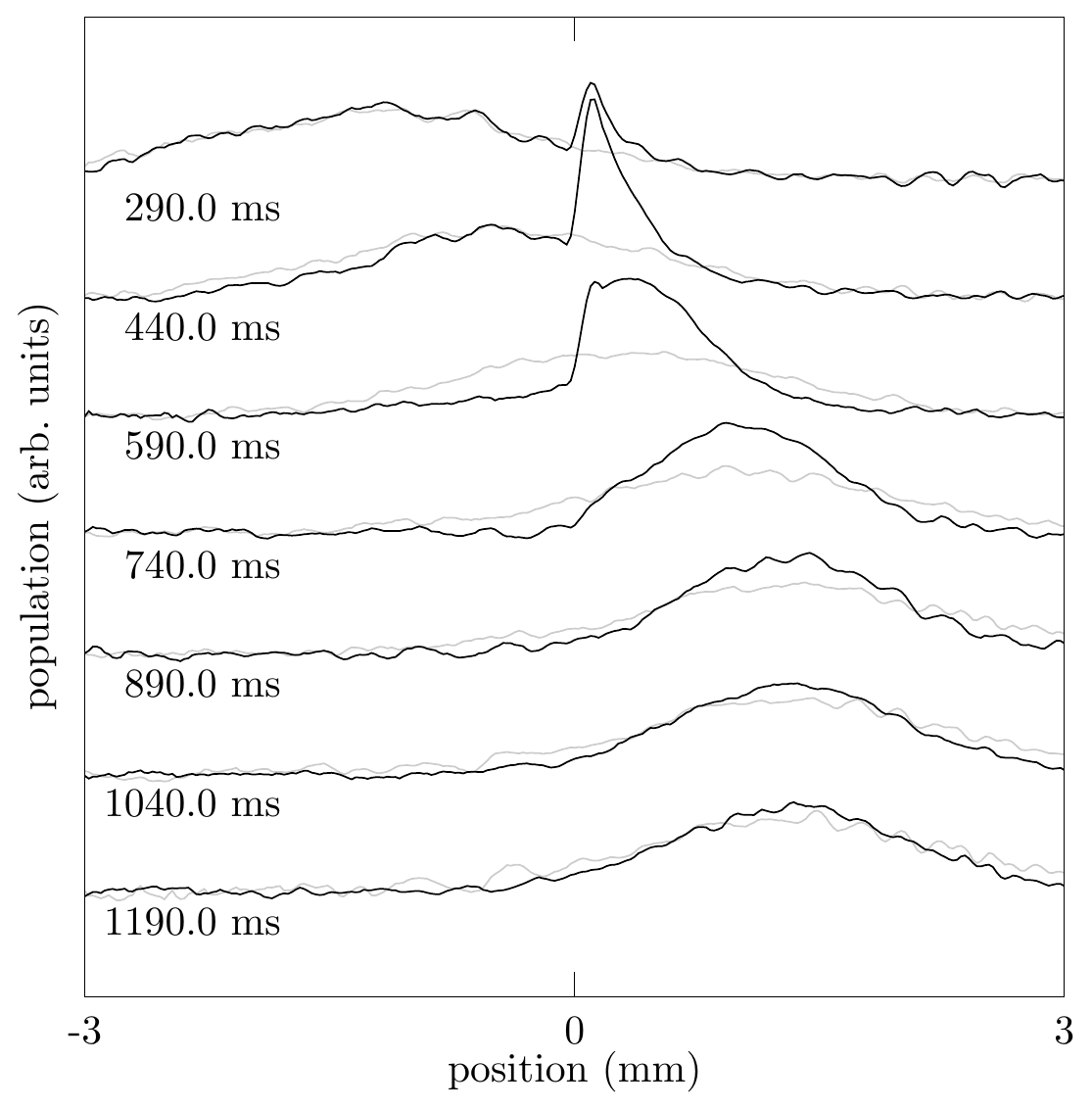}
\caption{Spatial distributions of atoms in the dipole trap with (black curves) and without (gray curves) the one-way barrier.  The barrier beams are located at the plot center.  Each curve is an average of $16$ repetitions of the experiment, smoothed slightly for clarity.\label{fig:waterfall}}
\end{figure}


The main results for the cooling scheme are presented in Fig.~\ref{fig:waterfall}, which overlays the atomic spatial distributions in the dipole trap in the presence and absence of the one-way barrier.  At time $t=290 \ \mathrm{ms}$, the hot atomic distributions are at the position where the hottest atoms start to encounter the barrier.  The data at $t=440 \ \mathrm{ms}$ and $590 \ \mathrm{ms}$ illustrate the barrier's effect while it is in contact with the atomic distribution.  The atoms trapped by the barrier near their turning points are visible as narrow peaks on the barrier's right hand side.  For the distributions with the barrier (in black) at these times, the width is not an appropriate indicator of temperature because the harmonic dipole trap potential is effectively cut in half by the one-way barrier potential.  The atoms trapped on the right-hand side of the barrier then follow it to the bottom of the potential, with minimal increase in their kinetic energy.  This is manifest in the narrower distributions at times $t=740 \ \mathrm{ms}$ and $890 \ \mathrm{ms}$, clearly illustrating the cooling effect of the moving barrier.  By time $t=1190 \ \mathrm{ms}$, the cooling effect of the barrier is less obvious because the cooled ensemble begins to heat up.  This slow increase in temperature is most likely due to a combination of dipole-trap intensity fluctuations, uncertainty in the overlap of the dipole-trap focus with the foci of the barrier beams, and anharmonic coupling to the transverse dimensions that were not cooled.

\begin{figure}
\includegraphics[width=\columnwidth]{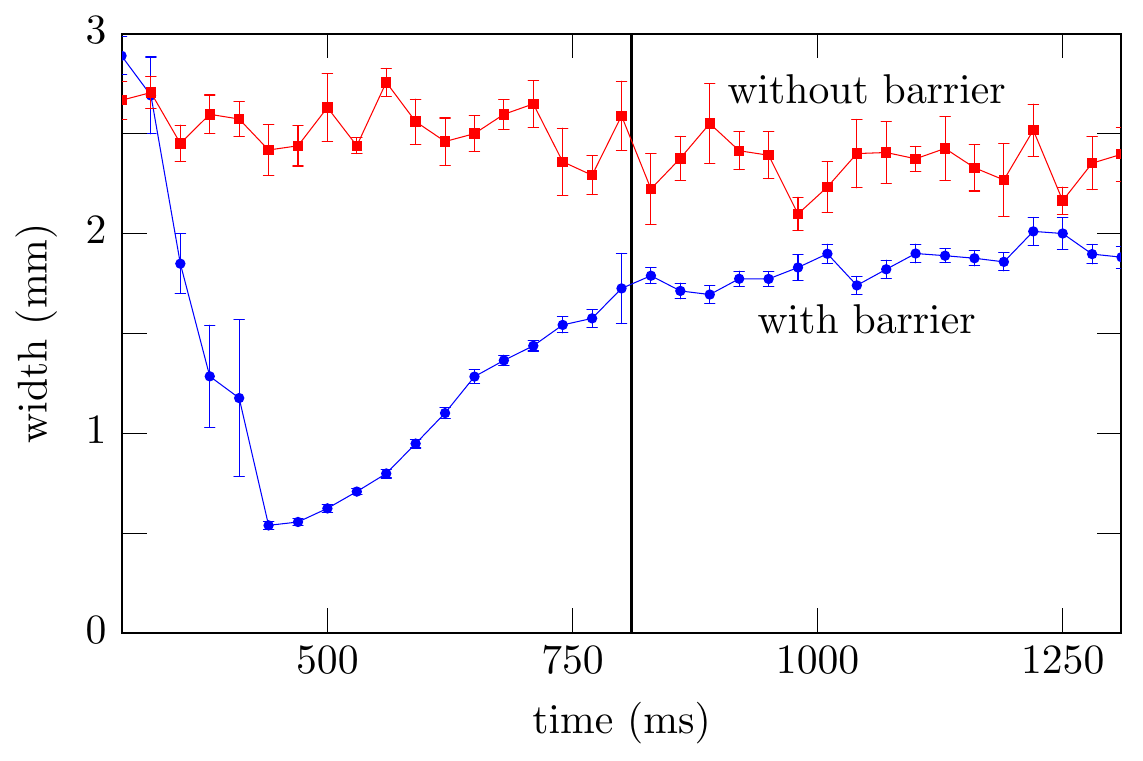}
\caption{(Color online.) The width of the atomic distribution in the dipole trap as function of time, measured from the start of the sweep, with (squares) and without (circles) the one-way barrier.  The vertical line indicates the time when the sweep has finished.  Each data point is an average of $16$ repetitions of the experiment, with the standard deviation as the error.\label{fig:widths}}
\end{figure}

The width of the atomic ensemble before, during, and after the cooling process is shown in Fig.~\ref{fig:widths}, where the vertical line delineates the end of the sweep.  As mentioned earlier, the width does not accurately reflect the temperature while the barrier interacts with atoms, as is the case for the data with the barrier in Fig.~\ref{fig:widths} prior to the end of the translation.  Cooling in one dimension, we reduce the temperature of the atoms by a factor of $1.7$, cooling the sample to about $100 \ \mathrm{\mu K}$ after completing the sweep.  Including spatial compression, this corresponds to a factor of $1.7$ reduction in the phase-space volume.  The increase in phase-space density is less than has been reported for an earlier one-way-barrier scheme \cite{bannerman2009}, however, we implemented an all-optical realization, necessitating higher powers and more scattering, while cooling nearly all of the atoms in the initial sample.  Comparison with simulations reveals that a factor of $9$ cooling ought to be achievable with our system, a discrepancy we attribute to the previously described heating effects counteracting the cooling process.  As the cooled atoms sit in the dipole trap, their temperature begins to increase, the same phenomenon observed in the time $t=1040 \ \mathrm{ms}$ and $t=1190 \ \mathrm{ms}$ atomic-distribution data presented in Fig~\ref{fig:waterfall}.

For the sweep to be adiabatic, the distance the barrier moves during one longitudinal oscillation period of the atoms must be small.  This ensures that all atoms are trapped near their turning points, when their kinetic energies are lowest.  The barrier must also be translated slowly enough so as not to significantly increase the kinetic energy of the trapped atoms as they ``fall'' toward the receding barrier.  The average speed of an atom during an oscillation period is approximately $100 \ \mathrm{mm/s}$, so we want to translate the trap at least an order of magnitude more slowly.  Treating the atoms classically and assuming a linear potential, the temperature of an atom when it ``catches up'' with the barrier is proportional to $v_0^{\;2}$, where $v_0$ is the sweep speed.  This implies that the effectiveness of the cooling scheme will continue to improve as the speed of the sweep is reduced.  

\begin{figure}
\includegraphics[width=\columnwidth]{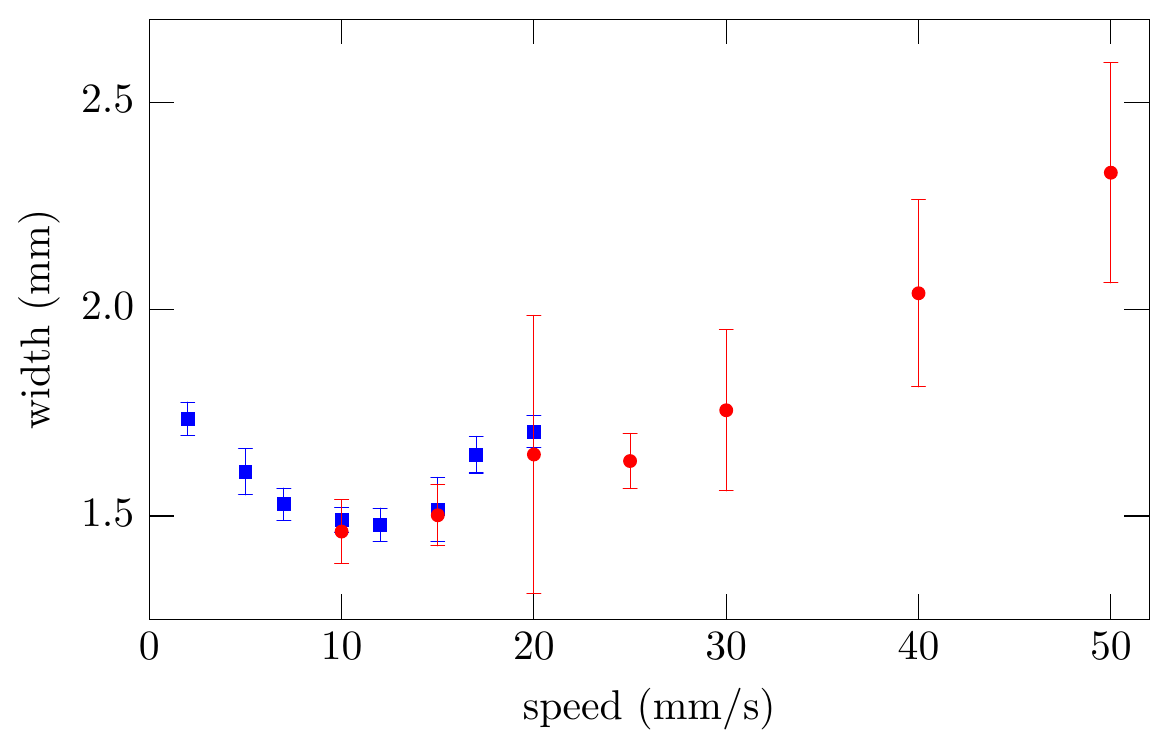}
\caption{(Color online.) The width of the atomic distribution after passing through the barrier, as function of translation velocity.  Square points represent data taken while the translation stage was coming to rest at $1.5 \ \mathrm{mm}$, while circular points represent data taken as the stage passed $1.5 \ \mathrm{mm}$ at constant velocity.  Each square (circular) data point represents an average of $16$ ($6$) repetitions of the experiment.\label{fig:velocity}}
\end{figure}

Figure~\ref{fig:velocity} shows the width of the atoms for a variety of different translation speeds.  At high speeds the deceleration times are significant, so the images were taken at the time the translation stage passed $1.5 \ \mathrm{mm}$ at constant velocity.  These data are represented by the circular points.  The speeds represented by the square points were slow enough that the images were taken as the stage decelerated to rest at a position of $1.5 \ \mathrm{mm}$.  For these images, the length of time after passing through the barrier coincided as closely as possible to the constant-velocity data for each velocity.  Neither the deceleration of the translation stage nor the variation in the timing of the data collection affects the width of the atomic distributions, as illustrated by the good agreement in the region of overlap in the data.  Virtually no velocity dependence is discernible until  $30 \ \mathrm{mm/s}$, a direct result of the heating effects described earlier that also limit our overall reduction in temperature.  We chose $5 \ \mathrm{mm/s}$ for the detailed measurement in Figs.~\ref{fig:waterfall} and \ref{fig:widths} because it provided a balance between improved cooling and atom loss due to scattered light from the barrier beam, a limitation of our system that we now discuss in more detail.

There are several factors limiting the effectiveness of the cooling scheme related to our particular implementation.  First, using a single-beam, far-detuned dipole trap allows us to treat the system as effectively one-dimensional.  However, a significant amount of energy can be present in the transverse dimensions \cite{thorn2009}.  Our work is limited by the fact we only cool the atomic sample in the longitudinal dimension.

The second significant limitation inherent in our experiment is the $6.8 \ \mathrm{GHz}$ hyperfine ground-state splitting in $^{87}$Rb, which restricts the detuning of the main barrier beam to small values.  The functionality of the one-way barrier is sensitive to scattering events that occur when an atom interacts with the main barrier beam during transmission and reflection.  Fewer scattering events reduce heating and trap loss in the presence of the barrier, resulting in longer trap lifetimes \cite{thorn2008,thorn2009}.  Trap lifetimes range from $700$ to $900 \ \mathrm{ms}$ when atoms are trapped against the barrier, measured after $200 \ \mathrm{ms}$, once the hottest atoms have been lost. The small detunings afforded by the $6.8 \ \mathrm{GHz}$ ground-state splitting produce around three scattering events during transmission and around four scattering events during reflection for the moving barrier.  This has serious implications for cooling with the one-way barrier, which requires repeated reflections off the barrier as it is adiabatically swept through the trap.  In particular, this prevents us from translating the barrier arbitrarily slowly because the lifetimes are small, and at some point the heating and trap losses will overcome any cooling benefit due to the slower sweep speed.

An obvious improvement to this cooling scheme is to increase the detuning of the main barrier beam from the transmitting and reflecting states.  One possibility for $^{87}$Rb is to exploit the $F=1$ magnetic sub-levels, recognizing that at a wavelength of $792.5 \ \mathrm{nm}$, the optical dipole potential vanishes for the $m_F=+1$ state, while it is positive for the other two sub-levels.  Thus, the transmitting state experiences no potential well or barrier, while the reflecting state has a detuning of $2.5 \ \mathrm{nm}$.  Another possibility is an implementation with a different atom or molecule with a more accommodating level structure.  We have worked out a specific example for $^{88}$Sr previously \cite{thorn2009}.  Clearly, the detuning of the barrier merits careful consideration when applying this cooling technique to new atomic and molecular species.

In summary, we have implemented a novel cooling scheme for $^{87}$Rb atoms using a moving, all-optical one-way barrier, demonstrated its effectiveness at cooling a sample of atoms, and addressed experimental limitations and improvements.



%




\begin{acknowledgments}
This research was supported by the National Science Foundation,
under project PHY-0547926.  We would like to thank Joseph Thywissen for his suggestion of using the fine structure of $^{87}$Rb to implement an improved one-way barrier.
\end{acknowledgments}

\providecommand{\noopsort}[1]{}\providecommand{\singleletter}[1]{#1}%

\end{document}